\documentclass{emulateapj}

\slugcomment{Astronomical Journal}
\shorttitle{HCN/HCO+ in ULIRGs}
\shortauthors{Imanishi et al.}

\begin{document}

\title{Millimeter Interferometric Investigations of the Energy Sources
of Three Ultraluminous Infrared Galaxies, UGC 5101, Mrk 273, and IRAS
17208$-$0014, based on HCN to HCO$^{+}$ Ratios} 

\author{Masatoshi Imanishi}
\affil{National Astronomical Observatory, 2-21-1, Osawa, Mitaka, Tokyo
181-8588, Japan} 
\email{masa.imanishi@nao.ac.jp} 

\author{Kouichiro Nakanishi}
\affil{Nobeyama Radio Observatory, Minamimaki, Minamisaku, Nagano,
384-1305, Japan} 
\email{nakanisi@nro.nao.ac.jp}

\and

\author{Kotaro Kohno}
\affil{Institute of Astronomy, University of Tokyo, 2-21-1, Osawa, Mitaka, 
Tokyo, 181-0015, Japan}
\email{kkohno@ioa.s.u-tokyo.ac.jp}

\begin{abstract}
We present interferometric observations of three ultraluminous
infrared galaxies (ULIRGs; UGC 5101, Mrk 273, and IRAS 17208$-$0014) in
the 3-mm wavelength range, using the Nobeyama Millimeter Array.  Both
the HCN (J=1--0) and HCO$^{+}$ (J=1--0) molecular lines were
observed simultaneously.  HCN emission was clearly
detected at the nuclear positions of these ULIRGs, and HCO$^{+}$
emission was detected at the nuclear positions of UGC 5101 and IRAS
17208$-$0014.  The HCN to HCO$^{+}$ brightness-temperature 
ratios toward the nuclei of the three ULIRGs were derived and compared
with those of lower luminosity galaxies known to be dominated by
active galactic nuclei (AGNs) or starbursts.  
In UGC 5101 and Mrk 273, where there is evidence for obscured AGNs from 
previous observations at other wavelengths, we found high 
HCN/HCO$^{+}$ ratios ($>$1.8) that are in the range found for 
AGN-dominated galaxies.  In IRAS 17208$-$0014, where the presence of 
a powerful obscured AGN has been unclear, the ratio ($\sim$1.7) is in
between the observed values for starburst- and AGN-dominated galaxies.   
The high HCN/HCO$^{+}$ brightness-temperature ratios in UGC 5101 and Mrk
273 could be the consequence of an HCN abundance enhancement, which is
expected from chemical effects of the central X-ray emitting AGN on the
surrounding dense molecular gas.  
Our proposed millimeter interferometric method based on HCN/HCO$^{+}$
ratios may be an effective tool for unveiling elusive buried AGNs at the
cores of ULIRGs, especially because of the negligible dust extinction at
these wavelengths.
\end{abstract}


\section{Introduction}

Ultraluminous infrared galaxies (ULIRGs), discovered during the {\it
IRAS} all-sky survey, radiate the bulk of their very large
luminosities ($L> 10^{12}L_{\odot}$) as dust emission in the infrared
\citep{sam96}.  Hence, powerful energy sources, such as starbursts or
active galactic nuclei (AGNs), must be present hidden behind dust.  
Since distant ULIRGs dominate the cosmic infrared background
emission, the sum of the dust-obscured activity in the universe
\citep{bla99}, establishing the energy sources of ULIRGs is
closely related to understanding the connections between obscured AGNs
and starbursts.  Distant ULIRGs are too faint to investigate in detail
with existing facilities and so the study of nearby ULIRGs continues
to be an important means of understanding the distant ULIRG
population.

It is now generally accepted that the majority of nearby ULIRGs are
energetically dominated by compact ($<$500 pc), highly dust-obscured
nuclear cores, rather than by extended ($>$kpc), weakly obscured
(A$_{\rm V}$ $<$ 15 mag) starburst activity in host galaxies
\citep{soi00}. Therefore, a primary goal in the study of nearby ULIRGs
is to establish whether the obscured compact cores are very compact
starbursts or AGNs.  
Optical spectroscopy is an often used way to distinguish an AGN from a
starburst.  In the
presence of an AGN obscured by dust in a torus geometry,
clouds that are photo-ionized by the radiation from the AGN develop
along the torus axis above a torus scale height (the 
so-called narrow line regions: NLRs) \citep{ant93}.
The NLRs produce optical emission lines whose flux
ratios differ from the clouds in starburst galaxies;
galaxies that contain an AGN obscured by torus-shaped dust are
classified optically as Seyfert 2s \citep{vei87}.  Hence, this type of
obscured AGN can be identified by using optical spectroscopy.  
However, the nuclear regions of ULIRGs contain sufficiently highly
concentrated molecular gas and dust that putative AGNs may be obscured
by dust in virtually all directions \citep{sam96}.  Such {\it buried}
AGNs 
%-----------
\footnote{
We use the term ``obscured AGN'' if the AGN is obscured along
our line of sight, and ``buried AGN'' if the AGN is obscured along
virtually all sightlines.
}  
%-----------
without significant NLRs are classified optically as
``non-Seyferts'', and so very difficult to find \citep{mai03}; however,
they are predicted to be abundant in the universe \citep{fab02}.  In
fact, the initial evidence of powerful buried AGNs has been detected in
several optical non-Seyfert ULIRGs 
\citep{dud97,soi00,imd00,tra01,idm01,imm03,ima03,idm06}, so that it is
essential to understand the contribution from buried AGNs to the total
energy output of the ULIRG population. 

One of the most effective ways to distinguish a buried AGN from a very
compact starbust as the primary energy source in the core of a ULIRG
is to investigate whether there is a strong X-ray-emitting source.  A
luminous buried AGN should exhibit intrinsically strong X-ray emission 
at E $>$ 2 keV from the vicinity of the accretion disk around the
central supermassive black hole \citep{elv94}, whereas in a starburst,
the intrinsic X-ray emission above $E\sim2$ keV, relative to bolometric
luminosity, is significantly weaker \citep{ran03}.  
Therefore, the detection of intrinsically 
strong, but highly absorbed, X-ray emission can provide strong evidence
for a powerful buried AGN.  Although moderately X-ray absorbed,
Compton thin (N$_{\rm H}$ $<$ 10$^{24}$ cm$^{-2}$) AGNs are easy to
find with currently available X-ray data at E = 2--10 keV
\citep{fra03,pta03,imt04}, it is very difficult to 
unveil and quantify the luminosities of Compton thick (N$_{\rm H}$ $>$
10$^{24}$ cm$^{-2}$) buried AGNs using such data.  Given that most of
the AGNs detected so far in optical Seyfert-2 
ULIRGs \citep{fra00,iwa01,bra03,iwa05} and non-Seyfert galaxies with
L$_{\rm IR}$ $\lessapprox$ 10$^{12}$L$_{\odot}$
\citep{don96,vig99,del02} suffer from Compton-thick absorption, it is
very likely that the majority of the putative buried AGNs in
optical non-Seyfert ULIRGs are also Compton thick.  High-sensitivity
observations at E $>$ 10 keV using a next-generation X-ray satellite
are required to properly evaluate the energetic importance of buried
AGNs in a large ULIRG sample.

Another approach to investigating whether a ULIRG core has an
intrinsically strong X-ray emitting source is to search for signatures
of the effects of the luminous X-ray source on the surrounding
interstellar medium.  Around a buried AGN, X-ray dissociation regions
(XDRs; Maloney et al. 1996) should develop, rather than the
photo-dissociation regions (PDRs) usually found in starburst regions.
XDRs and PDRs can be distinguished if they show different flux ratios
of emission lines.  In particular, emission lines in the far-infrared
to millimeter wavelength range are potentially a powerful tool,
because low dust extinction at these wavelengths makes it possible to
uncover the signatures of XDRs around powerful buried AGNs that are
usually deeply embedded in dust.  \citet{koh01,koh03} and
\citet{koh05} proposed that the HCN (J=1--0) ($\lambda$=3.38 mm),
HCO$^{+}$ (J=1--0) ($\lambda$=3.36 mm), and CO(J=1--0) ($\lambda$=2.60
mm) lines are useful for this purpose, and argued that AGN-dominated
galaxies (XDRs) show systematically larger HCN/HCO$^{+}$ and HCN/CO
ratios in brightness temperature than starburst-dominated galaxies
(PDRs).  \citet{ima04} pointed out that the HCN/HCO$^{+}$ ratio is
particularly effective for 
investigating the nature of heavily obscured ULIRG cores for the
following reasons: (1) since both molecular lines  probe high-density
molecular gas ($n_{\rm H_2}$ $>$10$^{4}$ cm$^{-3}$), their relative
ratios are not strongly contaminated by more extended lower density gas,
whose relative fraction can vary among different galaxies, and  
(2) the wavelengths of the two lines are sufficiently close that they
are observable simultaneously with wide-band receiving 
systems, reducing the net observing time and systematic errors.
Since HCN and HCO$^{+}$ are both moderately strong lines,
this method is applicable to a larger number of galaxies than that
proposed by \citet{use04}, which uses various faint molecular lines
for the purpose of distinguishing between XDRs and PDRs.

Interferometric maps are preferable to single-dish large-aperture
observations for investigating the cores of ULIRGs in the millimeter
wavelength range, because they reduce the contamination from extended
star-formation emission in the host galaxy. We applied 
this millimeter interferometric method to the infrared luminous galaxy
NGC 4418 (L$_{\rm IR}$ $\sim$ 10$^{11}$L$_{\odot}$), which likely
contains a powerful buried AGN \citep{dud97,spo01}, and were
successful in finding signatures of a buried AGN from the
HCN/HCO$^{+}$ ratio \citep{ima04}, demonstrating the potential of this
method. 
In this paper, we report the application of this interferometric
HCN/HCO$^{+}$ method to three ULIRGs: UGC 5101, Mrk 273, and IRAS
17208$-$0014.  Throughout this paper, we adopt $H_{0}$ $=$ 75 km
s$^{-1}$ Mpc$^{-1}$, $\Omega_{\rm M}$ = 0.3, and $\Omega_{\rm
\Lambda}$ = 0.7.

\section{Targets}

The target ULIRGs were selected based on the following criteria: 
1) Strong HCN emission is detected with single-dish radio telescopes, so
that the study of HCN and HCO$^{+}$ emission-line ratios at the nuclei
is feasible with the Nobeyama Millimeter Array (NMA), our observing
facility. 
(2) They are nearby ($z <$ 0.06), so that redshifted HCN and HCO$^{+}$
emission lines are observable with the NMA receiving systems.  The
three ULIRGs UGC 5101, Mrk 273, and IRAS 17208$-$0014 were selected
as the first targets.  Table 1 summarizes the detailed information about
these sources.  An angular scale of 1$''$ corresponds to a physical
size of 0.7--0.8 kpc for the three ULIRGs.

UGC 5101 (z = 0.040) is included in the original ULIRG sample
\citep{san88a} and is a well-studied ULIRG. It is classified optically 
as a LINER \citep{vei95} or as a Seyfert 2 in a slightly different
classification scheme by a different group \citep{gon99}.
\citet{idm01} and \citet{imm03} performed infrared 3--4 $\mu$m
spectroscopy and detected polycyclic aromatic hydrocarbon (PAH)
emission at 3.3 $\mu$m, a starburst indicator, as well as signatures
of a powerful AGN deeply buried at the core.  They inferred that the
infrared luminosity of UGC 5101 is dominated by the buried AGN, rather
than the detected starburst activity.  Radio VLBI observations
\citep{smi98,lon03} and a near-infrared high-spatial-resolution image
\citep{sco00} showed compact, high surface-brightness emission, which
was interpreted as an AGN by these authors.  Similar compact emission
was detected in the infrared at 12.5 $\mu$m \citep{soi00}.  The
presence of a powerful obscured AGN was confirmed by
subsequent 2--10 keV X-ray observations \citep{ima03}.  \citet{arm04}
detected weak high-excitation forbidden emission lines originating in
the NLRs in a mid-infrared 5--40 $\mu$m high-resolution spectrum.
Judging from the detection of NLR-originating emission lines in the
optical \citep{gon99} and mid-infrared \citep{arm04}, there may be
a small angular range that is transparent to the AGNs' ionizing photons 
(mostly UV and soft X-rays).  HCN (J=1--0) data based on a
single-dish telescope are available \citep{gao04} and indicate that if
a significant fraction of the detected HCN emission comes from the
nucleus of UGC 5101, then an investigation of the HCN/HCO$^{+}$ ratio
at the core should be feasible using the NMA with a reasonable
integration time.

Mrk 273 (z = 0.038) is also a well-studied ULIRG \citep{san88a}. 
It is classified optically as a Seyfert 2 \citep{vei99}.  Its Seyfert-2
classification and the detection of strong NLR-originating
high-excitation forbidden emission lines in the mid-infrared 5--11
$\mu$m spectrum \citep{gen98} suggest that the dust distribution
around the putative AGN is torus-shaped and that a significant amount
of ionizing radiation is escaping the AGN.  2--10 keV X-ray data
have revealed the presence of a moderately absorbed Compton-thin AGN
\citep{iwa99,xia02,bal05}.  The presence of an obscured AGN in Mrk 273
is also suggested from (1) a smaller equivalent width of 3.3 $\mu$m
PAH emission than the typical value for starburst galaxies
\citep{imd00}, (2) deep 9.7 $\mu$m silicate dust absorption and weak
PAH emission features in the 8--13 $\mu$m spectrum
\citep{dud99,soi02}, and (3) compact high surface brightness
mid-infrared emission at 12.5 $\mu$m \citep{soi00}.  The HCN flux from
Mrk 273 observed with a single-dish telescope is as strong as that
from UGC 5101 \citep{gao04}, so that investigations of nuclear HCN and
HCO$^{+}$ emission are also feasible for Mrk 273.

IRAS 17208$-$0014 (z = 0.042) is classified optically as an HII-region
\citep{vei95}.  Unlike other ULIRGs, the mid-infrared 8--23 $\mu$m
emission of IRAS 17208$-$0014 is spatially extended, with no prominent
core \citep{soi00}.  Its radio emission as detected with the VLA is
relatively compact \citep{mar89,con96}, but its surface brightness is
not high enough to indicate strongly the presence of an AGN
\citep{dia99}.  
Its 8--13 $\mu$m spectrum is typical of starburst galaxies, with
prominent PAH emission features \citep{dud99}
Its 3--4 $\mu$m spectrum also displays a clear 3.3 $\mu$m PAH emission
feature, but the detection of a moderately strong 3.1 $\mu$m ice
absorption feature suggests an energy source more centrally-concentrated
than the surrounding dust, such as a buried AGN \citep{ris06,ima06}. 
However, in IRAS 17208$-$0014, the AGN signature is much weaker than 
UGC 5101 and Mrk 273.  
The detected HCN flux using a single-dish telescope is the largest among
the three ULIRGs \citep{gao04}.

\section{Observations and Data Reduction}

Interferometric observations of HCN (J=1--0) ($\lambda_{\rm rest}$ =
3.3848 mm or $\nu_{\rm rest}$ = 88.632 GHz) and HCO$^{+}$ (J=1--0)
($\lambda_{\rm rest}$ = 3.3637 mm or $\nu_{\rm rest}$ = 89.188 GHz)
were performed with the NMA and RAINBOW Interferometer at Nobeyama Radio
Observatory (NRO) between December 2004 and March 2005.  Table 2
summarizes the detailed observing log.  The NMA consists of six 10-m
antennas, and observations were performed using the D (the longest baseline
was 82 m), C (163 m), and AB (351 m) configurations.  The RAINBOW
interferometer is a seven-element combined array that includes six
10-m antennas (NMA) and the NRO 45-m telescope.  The RAINBOW
observations were performed when the NMA array was in the AB 
configuration, and the longest baseline was 410 m.  The sensitivity of
RAINBOW is about twice that of the NMA array in the 3-mm wavelength
range, because the inclusion of the NRO 45-m telescope increases the
total aperture size by a factor of four.

The backend was the Ultra-Wide-Band Correlator (UWBC) \citep{oku00},
which was configured to cover 1024 MHz with 128 channels at 8-MHz
resolution.  The central frequency was set to 85.49, 85.67, and
85.24 GHz for UGC 5101, Mrk 273, and IRAS 17208$-$0014, respectively,
so that both the redshifted HCN and HCO$^{+}$ lines were
 observable simultaneously.  A bandwidth of 1024 MHz corresponds to
$\sim$3500 km s$^{-1}$ at $\nu$ $\sim$ 86 GHz.  The field of view at
this frequency is $\sim$26$''$ (full width at half maximum; FWHM) and
$\sim$77$''$ (FWHM) for RAINBOW and NMA, respectively.  Since the
Hanning window function was applied to reduce side lobes in the spectra,
the effective resolution was widened to 16 MHz or 55 km s$^{-1}$ at
$\nu$ $\sim$ 86 GHz.  To calibrate the passband across the 128 channels,
the bright quasars 3C 84 (UGC 5101), 0355+508 (Mrk 273), and 3C 454.3
and 3C279 (IRAS 17208$-$0014) were observed every observing day.
Quasars 0954+658, 1418+546, and 1741$-$038 were used to calibrate
temporal variation in the visibility amplitude and phase for UGC
5101, Mrk 273, and IRAS 17208$-$0014, respectively.

The UVPROC-II package developed at NRO \citep{tsu97} and the AIPS
package of the National Radio Astronomy Observatory were used
for standard data reduction.  Corrections for the antenna baselines,
bandpass properties, and the time variation of the visibility amplitude
and phase were applied to all of the data.  A fraction of the data
with large phase scatter due to bad radio seeing was removed from our
analysis.  After clipping a small fraction of data of unusually high
amplitude, the data were Fourier-transformed using a natural {\it UV}
weighting.  The flux calibration was made using the observations of
the appropriate quasars, 1749+096 (UGC 5101), 1741$-$038 and 0355+508
(Mrk 273), or 1741$-$038 (IRAS 17208$-$0014), whose flux levels
relative to Uranus or Neptune had been measured at least every month 
between December 2004 and March 2005.  A conventional
CLEAN method was applied to deconvolve the synthesized beam pattern.
The total net on-source integration times were $\sim$29 hours (UGC
5101), $\sim$8 hours (Mrk 273), and $\sim$18 hours (IRAS
17208$-$0014).  The synthesized beam sizes are 5$\farcs$8 $\times$
4$\farcs$5 (position angle is 45$^{\circ}$ west of north; PA =
$-$45$^{\circ}$) for UGC 5101, 1$\farcs$9 $\times$ 1$\farcs$5 (PA =
$-$66$^{\circ}$) for Mrk 273, and 3$\farcs$1 $\times$ 2$\farcs$4 (PA =
$-$1$^{\circ}$) for IRAS 17208$-$0014. 
Absolute positional uncertainties of NMA/RAINBOW maps were estimated to 
be much less than 1 arcsec.

\section{Results}

HCN (J=1--0) or HCO$^{+}$ (J=1--0) data are available for several
ULIRGs \citep{sol92,cur00,gao04}, but observations have generally been
performed with single-dish telescopes.  \citet{rad91} showed an
interferometric HCN map of the ultraluminous infrared galaxy Arp 220.
With this exception, no detailed morphological
information for HCN emission exists for ULIRGs.  Therefore, this work
allows us to reveal for the first time the spatial distribution of HCN
(J=1--0) in the ULIRG population other than Arp 220. Ours are also the
first published interferometric maps of ULIRGs for HCO$^{+}$ (J=1--0)
emission. 

Figure 1 shows the contours of the continuum emission from UGC
5101, Mrk 273, and IRAS 17208$-$0014.  Signals away from the HCN and
HCO$^{+}$ emission lines were combined to produce the continuum maps.
Strong continuum emission is found in all ULIRGs and the continuum peaks
are spatially coincident with the nuclear positions of these ULIRGs.
Table 3 summarizes the estimated continuum flux levels.
The detection significance is $>6\sigma$ (UGC 5101), $>4\sigma$ (Mrk
273), and $>10\sigma$ (IRAS 17208$-$0014). 
The spatial extents of the continuum emission agree with the beam sizes
within $\lesssim$50\%. 

After the continuum emission was subtracted, we investigated the
spatial distribution of the HCN and HCO$^{+}$ emission.  Figure 2 shows
the integrated intensity maps of the HCN emission for UGC 5101, Mrk 273,
and IRAS 17208$-$0014.  Figure 3 displays the HCO$^{+}$ integrated
intensity maps for UGC 5101 and IRAS 17208$-$0014.  
HCO$^{+}$ emission is not clearly detected from the nucleus of Mrk 273
in our interferometric map ($<3\sigma$).
With this exception, the HCN and HCO$^{+}$ emission lines are clearly
($>10\sigma$) detected.   
The detected emission lines are coincident with the nuclear positions of
the ULIRGs and has spatial extents consistent with
the beam sizes within $\lesssim$50\%. 
For UGC 5101 and IRAS 17208$-$0014, the signal-to-noise ratios around
the HCN and HCO$^{+}$ lines are good enough to present channel maps,
which are shown in Figures 4 and 5, respectively.

Figure 6 shows continuum-subtracted nuclear spectra around the HCN and
HCO$^{+}$ emission lines.  The presence of HCO$^{+}$ emission from Mrk
273 is unclear, but HCN emission from all the ULIRGs, and HCO$^{+}$
emission from UGC 5101 and IRAS 17208$-$0014 are clearly recognizable.
Figure 7 presents the Gaussian fits to these detected emission lines.
The central velocity and line width of the Gaussian were determined
independently for the HCN and HCO$^{+}$ lines.  The HCN emission from
UGC 5101 suggests the presence of double peaks, which may be related
to the nuclear rotating molecular disk probed by the CO (J=1--0) line
\citep{gen98}.  Therefore, a double Gaussian fit was also attempted for
the HCN line of UGC 5101. 
For the HCN and HCO$^{+}$ lines from IRAS
17208$-$0014, we added a constant continuum to subtly
adjust the residual continuum level in the vicinity of the HCN
and HCO$^{+}$ lines.  Table 4 summarizes the Gaussian fitting results.

The local standard of rest (LSR) peak velocities of HCN emission 
from Mrk 273 (11310 km s$^{-1}$) and IRAS 17208$-$0014 (12860 km s$^{-1}$) 
agree, within 25 km s$^{-1}$, with those of CO peak velocities 
(11324 km s$^{-1}$ for Mrk 273 and 12837 km s$^{-1}$ for IRAS
17208$-$0014; Downes \& Solomon 1998). 
HCN and CO \citep{dow98} line profiles are also similar for the nuclei
of Mrk 273 and IRAS 17208$-$0014. 
For the double Gaussian fit of HCN emission from UGC 5101, the velocity 
difference of the two components is $\sim$400 km $^{-1}$, which is
roughly comparable to that of CO ($\sim$480 km s$^{-1}$; Genzel et
al. 1998). 

The integrated fluxes of HCN and HCO$^{+}$ at the peak position,
estimated from the Gaussian fits, are summarized in Table 5.  For the
HCN emission from UGC 5101, the single and double Gaussian fits give
fluxes that are consistent to within $<$5\%.  Therefore, we adopt the
values based on single Gaussian fits for all sources.  The derived HCN
fluxes correspond to $\sim$60\% (UGC 5101), $\sim$30\% 
%-----
\footnote{
This value seems low, but we have no clear answer. 
The flux difference between the single-dish and interferometric
observations may be due to discrepancies in the absolute flux scale
between the two observations, or the presence of extended HCN emission
unrecovered by our interferometric observations (less likely), or
possibly a substantial overestimate of the HCN flux by Gao \& Solomon
(2004), coming from relatively large scatters in their single-dish data
(their Figure 2). 
}
%-----
(Mrk 273), and
$\sim$55\% (IRAS 17208$-$0014) of the HCN flux measurements from
single-dish radio telescopes \citep{gao04}.

The HCN/HCO$^{+}$ ratios in brightness temperature ($\propto$
$\lambda^{2}$ $\times$ flux density) are also summarized in Table 5.
For the purpose of detecting the effects of putative AGNs on the
surrounding interstellar medium at ULIRG cores, we regard our
HCN/HCO$^{+}$ ratios as a more appropriate tool than the ratios
previously derived from single-dish telescope observations
\citep{ngu92,sol92}, because (1) we can reduce the contamination from
the extended flux of the host galaxy, and 
(2) in our NMA data, the HCN and HCO$^{+}$ lines were observed
simultaneously with the same receiver and same correlator unit, so that
possible systematic errors should be minimal.  Although the {\it
absolute} flux calibration uncertainty of the NMA and RAINBOW data might
be as large as $\sim$20\%, this uncertainty does not affect the
HCN/HCO$^{+}$ ratios.  The uncertainty in the ratios is dominated by
statistical noise and fitting errors (see Figure 7).

\section{Discussion}

\subsection{Comparison of HCN/HCO$^{+}$ Ratios with Other Galaxies}

Figure 8 shows the HCN/HCO$^{+}$ and HCN/CO ratios in brightness
temperature for UGC 5101, Mrk 273, and IRAS 17208$-$0014, together
with previously obtained data for nearby starburst and Seyfert
galaxies \citep{koh01,koh05}, and the AGN-powered infrared luminous
galaxy NGC 4418 \citep{ima04}.  The ratios for the three main nuclei
of the infrared luminous merging galaxy Arp 299 (Imanishi et al. 2006,
in preparation) are also plotted.  \citet{koh01} and \citet{koh05}
found empirically that AGN-dominated galaxies tend to have higher
HCN/HCO$^{+}$ and HCN/CO ratios than starburst-dominated galaxies:
AGN-dominated galaxies should be distributed on the upper right-hand
side, whereas starburst-dominated galaxies should occupy the lower
left region in Figure 8.

The HCN/CO ratio in {\it brightness temperature} (abscissa) may
increase if the HCN {\it abundance} relative to CO is enhanced by
X-ray radiation from an AGN \citep{lep96}.  However, another possible
explanation for the enhanced HCN/CO brightness-temperature
ratio is an increase in molecular gas density. Since the dipole
moment of HCN ($\mu$ $\sim$ 3 debye; Millar et al. 1997) is much larger than
that of CO ($\mu$ $\sim$ 0.1 debye; Millar et al. 1997), HCN emission traces
a much higher density molecular gas ($n_{\rm H_2}$ $>$ 10$^{4}$
cm$^{-3}$) than is traced by CO ($n_{\rm H_2}$ $\sim$ 10$^{2-3}$
cm$^{-3}$).  The HCN/CO brightness-temperature ratio can increase if a
galaxy contains a larger fraction of high-density molecular gas
(e.g., Solomon et al. 1992; Helfer \& Blitz 1993; Kohno et al. 1999; 
Curran et al. 2000). 
In fact, ULIRGs are expected to have a larger fraction of high-density 
molecular gas than less infrared-luminous galaxies do \citep{sam96}, so
that HCN/CO brightness-temperature ratios can be high in ULIRGs,
even without the HCN abundance enhancement caused by powerful AGNs.
Furthermore, the HCN and CO data were taken on different observing
runs, with different beam sizes, under different weather conditions,
all of which could introduce additional systematic uncertainties.
Therefore, the interpretation of HCN/CO brightness-temperature ratios
in ULIRGs is not simple.

Such uncertainties are much reduced if we use the HCN/HCO$^{+}$
ratios.  Both HCN and HCO$^{+}$ ($\mu$ $\sim$4 debye; Botschwina et
al. 1993) probe high-density molecular gas ($n_{\rm H_2}$ $>$ 10$^{4}$
cm$^{-3}$), so that the HCN/HCO$^{+}$ brightness-temperature ratio is
insensitive to the different fraction of dense molecular gas, relative
to diffuse one, in individual galaxies.  Furthermore, since both the HCN
and HCO$^{+}$ emission lines were observed simultaneously with
NMA/RAINBOW for all the sources plotted in Figure 8, systematic errors
are not a serious concern. 
For this reason, we use the HCN/HCO$^{+}$ ratios (ordinate) in our
discussion of the possible presence of powerful AGNs.

Our new millimeter interferometric data for UGC 5101 and Mrk 273 reveal
that the HCN/HCO$^{+}$ brightness-temperature ratios are high and 
in the range expected for AGN-dominated galaxies (HCN/HCO$^{+}$
$\gtrsim$ 2; Kohno 2005).  This 
is consistent with the fact that previous data have found the
signatures of powerful AGNs in UGC 5101 and Mrk 273 ($\S$2).

The interpretation of our data for IRAS 17208$-$0014 is rather
unclear.  No strong evidence for a luminous buried AGN in IRAS
17208$-$0014 has been found at other wavelengths ($\S$2).  The
HCN/HCO$^{+}$ brightness-temperature ratio for IRAS 17208$-$0014
($\sim$1.7) is located in between the expected values for AGN- and
starburst-dominated galaxies.  
A powerful AGN possibly hidden in IRAS 17208$-$0014 may slightly
increase the HCN/HCO$^{+}$ ratio, but the presence of an AGN 
is not clear from our data.

\subsection{Interpretations of HCN/HCO$^{+}$ Ratios}

In ULIRGs, the profiles of HCN (J=1--0) and CO (J=1--0) lines are
generally similar (Solomon et al. 1992; Gao \& Solomon 2004; This
paper), despite the difference of their optical depths.  
The similar line profiles are naturally explained by the widely accepted
mist model \citep{sol87}, where a molecular gas cloud consists of a
large number of small clumps, with a filling factor of $<$1 at any fixed
velocity. 
In this model, even if each clump is optically thick for a 
molecular line, we can probe clumps at the background side of the
molecular gas cloud.

If each clump is optically thin for a molecular line, its flux 
will increase linearly with increasing abundance.   
However, each clump is thought to be optically thick for CO (J=1--0)
lines \citep{sol87,sco87}.
It is not clear whether each clump is optically thick or thin for 
HCN (J=1--0) and HCO$^{+}$ (J=1--0) lines, but we can roughly estimate
their optical depths ($\tau$), based on the formula of 
\begin{eqnarray}
\tau & \propto & A \times \lambda^{3} \times abundance,  
\end{eqnarray}
where {\it A} is the Einstein coefficient and is proportional 
to $\mu^{2}$, where $\mu$ is the value of dipole moment \citep{sco87}. 
If we adopt the HCN and HCO$^{+}$ abundance, relative to CO, are 
10$^{-4}$ $\sim$ 5$\times$10$^{-3}$ \citep{bla87}, the optical depths of
both HCN (J=1--0) and HCO$^{+}$ (J=1--0) lines are 0.2--2 times as large
as that of CO (J=1--0) line. 
Thus, this simple estimate suggests that each clump can be optically
thick for HCN (J=1--0) and HCO$^{+}$ (J=1--0) lines.
Nguyen-Q-Rieu (1992) also showed that each clump can have optical depth 
slightly larger than unity for HCN (J=1--0) line, if the volume filling 
factor of the clump is as small as f $\lesssim$ 0.1 \citep{sol87,kod05}.

For an optically thick case, although the relation between flux and
abundance is not so simple, their positive correlation is still expected for 
the following main reason. 
Assuming that each clump has a decreasing radial density profile
($\propto$ r$^{-1.5}$; Gierence et al. 1992), line emission comes mainly
from the layer where an optical depth reaches unity \citep{gie92}.
When a molecular abundance increases (decreases), the $\tau$ = 1 layer 
for that molecular line moves outward (inward).
Consequently, the surface area of the layer, or its area filling factor,
becomes larger (smaller), increasing (decreasing) the flux of that
molecular line.  
Unless the optical depth is extremely large, this effect can be
significant and flux will increase with increasing abundance, if not
completely linearly, as was demonstrated for the optically thick CO
(J=1--0) line \citep{ver95,wil95,sak96,ari96}.  
Hence, qualitatively, the high HCN/HCO$^{+}$ ratio in the brightness
temperature can be explained by an enhanced HCN/HCO$^{+}$
abundance ratio in dense molecular gas from which the bulk of the
observed HCN and HCO$^{+}$ fluxes originate.

Theoretical models allow the calculation of the abundance of
molecules, including HCN and HCO$^{+}$, as a function of distance from
a central UV illuminating source or X-ray emitting AGN deeply buried in
molecular gas and dust \citep{lep96,mei05}.  \citet{lep96} found that the
abundance of HCN can be enhanced relative to HCO$^{+}$ if a strong
X-ray source is present inside molecular gas.  \citet{mei05}
calculated the effects of a central X-ray-illuminating AGN on the
surrounding molecular gas.  They found that the HCN/HCO$^{+}$
abundance ratio in uniformly-distributed dense molecular gas (10$^{5.5}$
cm$^{-3}$), as probed by HCN and HCO$^{+}$, around a luminous
X-ray-emitting source can be substantially higher than the ratio in
dense molecular gas illuminated only by a central UV source (their
Fig.10, model 4). 
Thus, we have some theoretical basis for the enhanced HCN abundance,
relative to HCO$^{+}$, in the presence of an X-ray-emitting AGN.

However, further sophisticated model calculations are definitely
required to {\it quantitatively} interpret our observational results.  
Models available to date \citep{lep96,mei05} have only 
calculated the HCN/HCO$^{+}$ abundance ratio and showed that 
the abundance ratio is a function of distance from the central 
illuminating source.  
To convert from the HCN/HCO$^{+}$ abundance ratio to
brightness-temperature ratio, we have to know the region from where the
bulk of the observed HCN and HCO$^{+}$ fluxes come, which has not been
explicitly mentioned in any available model calculations.
Furthermore, the models assume uniform dust/gas distribution for
simplicity.  
If the distribution is clumpy, which is more realistic \citep{sol87}, 
the abundance ratio can change substantially.

In order for our millimeter interferometric HCN/HCO$^{+}$ method to
help to detect buried AGN signatures, high-density molecular
gas must be concentrated near the core region where putative AGNs reside
and can affect the gas.  If high-density molecular gas is distributed in
between double nuclei, for example as in NGC 6240 \citep{nak05}, then
our method may no longer work.  Furthermore, shocks may enhance
HCO$^{+}$ emission \citep{raw00,raw04}.  If shocks driven by outflows or
superwind from starbursts coexist with powerful buried AGNs, and if there is a
large amount of high-density molecular gas that is affected by the
shocks, then the HCN/HCO$^{+}$ brightness-temperature ratio may
decrease compared to that of a pure buried AGN.  
Thus, some fraction of the galaxies containing powerful AGNs can show
low HCN/HCO$^{+}$ ratios and may be missed when searching based on the
HCN/HCO$^{+}$ ratio. 

Despite these remained ambiguities, this millimeter
interferometric method can empirically be an effective means of
revealing the presence of deeply buried AGNs at ULIRG cores, because of
the negligible dust extinction in this wavelength range.  
If this method is established from a larger number of sources, 
future observations with high spatial resolution and high sensitivity
using ALMA are potentially a powerful probe for optically and even 2--10
keV X-ray-elusive buried AGNs in ULIRG cores.   

\section{Summary}

We performed millimeter interferometric observations of the HCN
(J=1--0) and HCO$^{+}$ (J=1--0) lines of the three ultraluminous
infrared galaxies, UGC 5101, Mrk 273, and IRAS 17208$-$0014, using 
NMA/RAINBOW.  These are the first interferometric maps of a ULIRG
sample for which the HCN (J = 1--0) and HCO$^{+}$ (J = 1--0) molecular
lines were observed simultaneously.  We extracted the HCN to HCO$^{+}$
brightness-temperature ratios at the cores of these ULIRGs, and
compared the ratios with those previously estimated for less-luminous,
AGN-dominated and starburst-dominated galaxies.  We then looked for
possible signatures of the influence of AGNs, strong X-ray emitters, on
the surrounding interstellar medium at the nuclei.  We found that:
\begin{enumerate}
\item Continuum and HCN emission were clearly detected in all ULIRGs.
      HCO$^{+}$ emission was found in UGC 5101 and IRAS 17208$-$0014.
      The spatial distribution of the emission, when detected,
      is coincident with the nuclear positions of these ULIRGs.
\item UGC 5101 and Mrk 273 showed high HCN/HCO$^{+}$ ratios in
      brightness temperature, as found in AGN-dominated galaxies.
      Previous observations had suggested luminous obscured AGNs in
      UGC 5101 and Mrk 273, and so our millimeter interferometric data
      may be revealing the signatures of the effects of a luminous 
      X-ray-emitting AGN on the surrounding interstellar medium at the
      nucleus. 
\item The HCN/HCO$^{+}$ ratio for IRAS 17208$-$0014 was in between the
      regions occupied by AGN- and starburst-dominated galaxies.  
      There was no clear evidence for a hidden AGN in IRAS 17208$-$0014
      based on previously obtained data at other wavelengths.  
      The presence of a luminous buried AGN in IRAS 17208$-$0014 remains
      uncertain based on currently available millimeter interferometric
      data. 
\item Although the sample size is still small, this millimeter
      interferometric HCN/HCO$^{+}$ method succeeded in confirming AGN
      signatures in the two ULIRGs (UGC 5101 and Mrk 273) that were
      previously known to possess luminous obscured AGNs.  Since dust
      extinction is negligible at this wavelength range, we argue that
      our method is potentially an important tool for unveiling powerful,
      but deeply buried, AGNs in optical ``non-Seyfert'' ULIRGs in the
      ALMA era. 
\end{enumerate}

\acknowledgments

We are grateful to NRO staff for their support during our
observing runs and to the anonymous referee for his/her valuable
comments. 
We thank S. Ishizuki, J. Koda, N. Z. Scoville, T. Sawada, and S. Takano
for useful discussions about the properties of molecular gas. 
We also thank S. K. Okumura for providing us detailed information
on the NMA.
M.I. was supported by a Grant-in-Aid for Scientific Research (16740117).
NRO is a branch of the National Astronomical Observatory, National
Institutes of Natural Sciences, Japan.
This research has made use of the SIMBAD database, operated at CDS,
Strasbourg, France, and of the NASA/IPAC Extragalactic Database
(NED), which is operated by the Jet Propulsion Laboratory, California
Institute of Technology, under contract with the National Aeronautics
and Space Administration.

\clearpage

%----------------- Tables -----------------%
%---- Table 1 ----%
\begin{deluxetable}{cccccccc}
\tabletypesize{\small}
\tablecaption{Detailed information about the observed ULIRGs
\label{tab1}}
\tablewidth{0pt}
\tablehead{
\colhead{Object} & \colhead{Redshift}   & 
\colhead{$f_{\rm 12}$}  & \colhead{$f_{\rm 25}$}  & 
\colhead{$f_{\rm 60}$}  & \colhead{$f_{\rm 100}$}  & 
\colhead{log $L_{\rm IR}$ (log $L_{\rm IR}$/$L_{\odot}$)} &
\colhead{Far-infrared} \\ 
\colhead{} & \colhead{}   & \colhead{(Jy)} & \colhead{(Jy)}  & \colhead{(Jy)} 
& \colhead{(Jy)}  & \colhead{(ergs s$^{-1}$)} & \colhead{Color} \\
\colhead{(1)} & \colhead{(2)} & \colhead{(3)} & \colhead{(4)} & 
\colhead{(5)} & \colhead{(6)} & \colhead{(7)} & \colhead{(8)}
}
\startdata
UGC 5101 & 0.040 & 0.25 & 1.03 & 11.54 & 20.23 & 45.5 (12.0)
& cool \\
Mrk 273  & 0.038 & 0.24 & 2.28 & 21.74 & 21.38 & 45.7 (12.1)
& cool \\ 
IRAS 17208$-$0014 & 0.042 & 0.20 & 1.66 & 31.14 & 34.90 & 45.9
(12.3) & cool \\
\enddata

\tablecomments{
Column (1): Object.
Column (2): Redshift.
Columns (3)--(6): f$_{12}$, f$_{25}$, f$_{60}$, and f$_{100}$ are
the {\it IRAS FSC}
fluxes at 12, 25, 60, and 100 $\mu$m, respectively.
Column (7): Decimal logarithm of the infrared (8$-$1000 $\mu$m) luminosity
in ergs s$^{-1}$ calculated as follows:
$L_{\rm IR} = 2.1 \times 10^{39} \times$ D(Mpc)$^{2}$
$\times$ (13.48 $\times$ $f_{12}$ + 5.16 $\times$ $f_{25}$ +
$2.58 \times f_{60} + f_{100}$) ergs s$^{-1}$
\citep{sam96}.
The values in parentheses are the decimal logarithms of the infrared
luminosities in units of solar luminosities.
Column (8): ULIRGs with f$_{12}$/f$_{25}$ $<$ ($>$) 0.2 are classified
as cool (warm) \citep{san88b}.
}
\end{deluxetable}

%---- Table 2----%
\begin{deluxetable}{lcl}
%\tabletypesize{\scriptsize}
\tablecaption{Observing log\label{tab2}}  
\tablewidth{0pt}
\tablehead{
\colhead{Object} & \colhead{Array} & \colhead{Observing Date} \\ 
\colhead{} & \colhead{Configuration} & \colhead{(UT)} 
%\colhead{(1)} & \colhead{(2)} & \colhead{(3)} & \colhead{(4)}
}
\startdata
UGC 5101 & D & 2004 Dec 22, 23, 24 \\
         & RAINBOW & 2005 Jan 20 \\
         & C & 2005 Mar 9, 10 \\ 
Mrk 273 & RAINBOW & 2005 Jan 22 and 31 \\
IRAS 17208$-$0014 & AB & 2005 Jan 26--28, Feb 7 \\
                  & C  & 2005 Mar 6, 7, 16--19, 26 \\
\enddata
\end{deluxetable}

%---- Table 3 ----%
\begin{deluxetable}{ccc}
%\tabletypesize{\scriptsize}
\tablecaption{Continuum emission \label{tab3}}  
\tablewidth{0pt}
\tablehead{
\colhead{Object} & \colhead{Flux} & \colhead{Frequency} \\ 
\colhead{} & \colhead{(mJy)} & \colhead{(GHz)} 
%\colhead{(1)} & \colhead{(2)} & \colhead{(3)} & \colhead{(4)}
}
\startdata
UGC 5101 & 6 & 85.5 \\
Mrk 273 & 6 & 85.6 \\
IRAS 17208$-$0014 & 9 & 85.3 \\
\enddata
\end{deluxetable}

%---- Table 4 ----%
\begin{deluxetable}{ccccc}
%\tabletypesize{\scriptsize}
\tablecaption{Gaussian fitting parameters of HCN and HCO$^{+}$ emission
lines \label{tab4}}   
\tablewidth{0pt}
\tablehead{
%\colhead{(1)} & \multicolumn{2}{c}{(2)} & \multicolumn{2}{c}{(3)} \\
\colhead{Object} & \multicolumn{2}{c}{LSR velocity} &
\multicolumn{2}{c}{FWHM}  \\   
\colhead{} & \multicolumn{2}{c}{(km s$^{-1}$)} & 
\multicolumn{2}{c}{(km s$^{-1}$)} \\ 
\colhead{} & \colhead{HCN} & \colhead{HCO$^{+}$} & \colhead{HCN} & 
\colhead{HCO$^{+}$}  \\
\colhead{(1)} & \colhead{(2)} & \colhead{(3)} & \colhead{(4)} &
\colhead{(5)} 
}
\startdata
UGC 5101 & 11830 & 11990 & 700 & 330  \\
         & 11585+11985 \tablenotemark{a} & 11990 
         & 270+375 \tablenotemark{a} & 330  \\
Mrk 273  & 11310 & \nodata & 450 & \nodata \\
IRAS 17208$-$0014 & 12860 & 12820 & 530 & 470 \\
\enddata

\tablecomments{
Col.(1): Object name.
Col.(2): LSR velocity \{v$_{\rm opt}$ $\equiv$ 
($\frac{\nu_0}{\nu}$ $-$ 1) $\times$ c\} of the
HCN emission peak.
Col.(3): LSR velocity of the HCO$^{+}$ emission peak.
Col.(4): Line width of the HCN emission at FWHM.
Col.(5): Line width of the HCO$^{+}$ emission at FWHM.
}

\tablenotetext{a}{
A double Gaussian fit of the HCN emission line (see $\S$4).}

\end{deluxetable}

%---- Table 5 ----%
\begin{deluxetable}{cccccc}
%\tabletypesize{\scriptsize}
\tablecaption{HCN, HCO$^{+}$, and CO emission \label{tab5}}   
\tablewidth{0pt}
\tablehead{
\colhead{Nucleus} & \colhead{HCN} & \colhead{HCO$^{+}$} &
\colhead{HCN/HCO$^{+}$} & \colhead{CO} & \colhead{Reference} \\   
\colhead{} & \colhead{(Jy km s$^{-1}$)} & \colhead{(Jy km s$^{-1}$)} & 
\colhead{} & \colhead{(Jy km s$^{-1}$)} & \colhead{} \\ 
\colhead{(1)} & \colhead{(2)} & \colhead{(3)} & \colhead{(4)} & 
\colhead{(5)} & \colhead{(6)}
}
\startdata
UGC 5101 & 6.6 & 2.7 & 2.5 & 50 & a \\
         & 3.6+2.7 \tablenotemark{a} & 
         & $>$2,$\sim$1 \tablenotemark{a} & &  \\
Mrk 273  & 4.8 & $<$2.7 & $>$1.8 & 78 & b \\
IRAS 17208$-$0014 & 14 & 8.1 & 1.7 & 132 & b \\
\enddata

\tablecomments{
Col.(1): Object name.
Col.(2): Integrated HCN intensity at the nucleus.
Col.(3): Integrated HCO$^{+}$ intensity at the nucleus.
Col.(4): HCN/HCO$^{+}$ ratio in brightness temperature 
($\propto$ $\lambda^{2}$ $\times$ flux density) at the nucleus.
The ratio is not affected by possible uncertainties in the absolute flux
calibration (see $\S$4).
Col.(5): Integrated CO intensity at the nucleus taken from the literature.
Col.(6): References for interferometric CO flux measurements.
(a): \citet{gen98}.; (b): \citet{dow98}.
}

\tablenotetext{a}{
Blue ({\it Left}) and red ({\it Right}) components for the 
double Gaussian fit (see $\S$4).
HCO$^{+}$ emission has a peak velocity close to that of the red HCN
component. 
We attribute most of the HCO$^{+}$ flux to the red component. 
}

\end{deluxetable}

%-----------------  Figures  ---------------% 

\clearpage

%---- Figure 1 ----%
\begin{figure}
\includegraphics[angle=0,scale=.8]{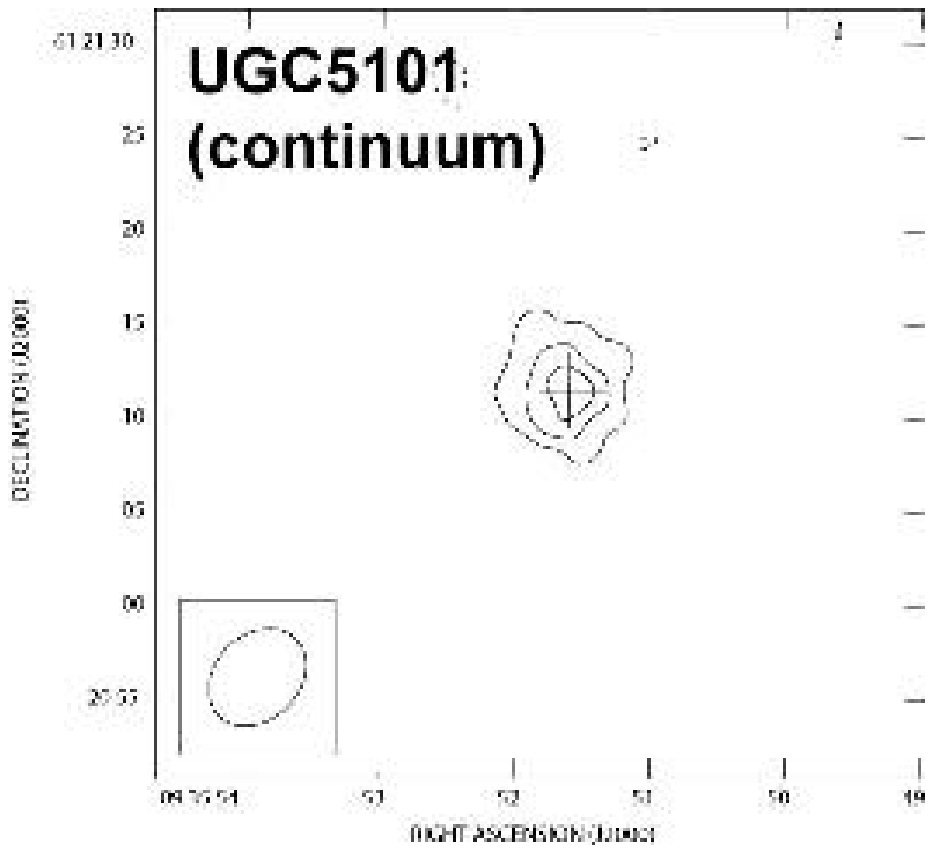} 
\includegraphics[angle=0,scale=.8]{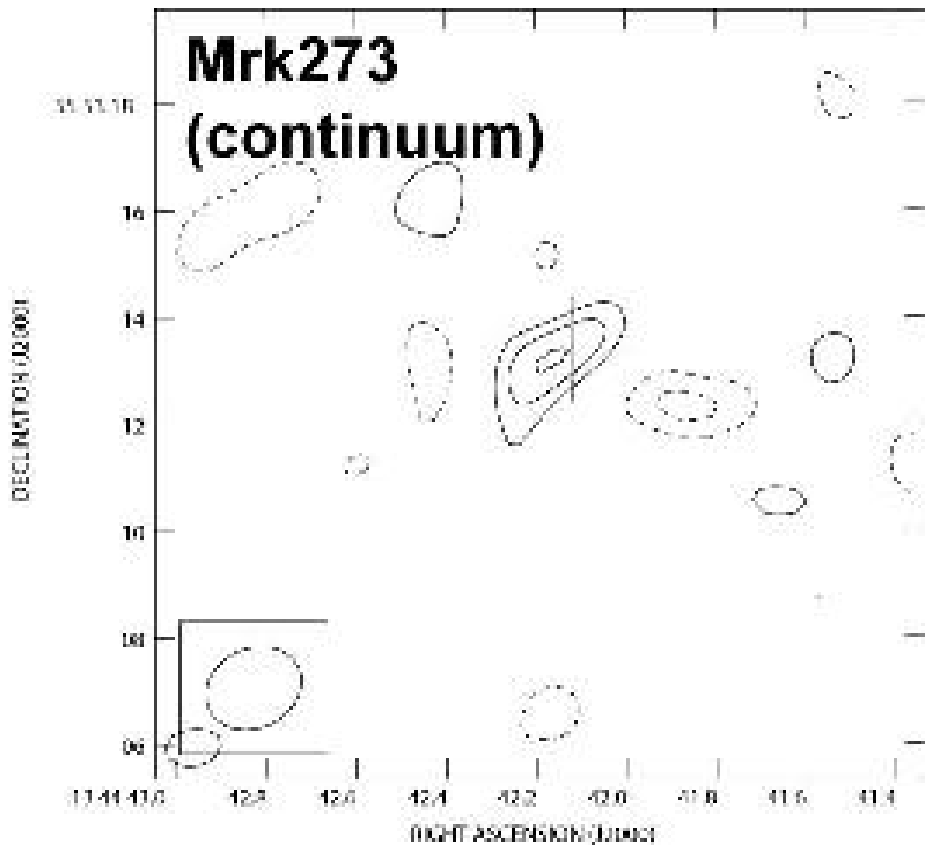} 
\includegraphics[angle=0,scale=.8]{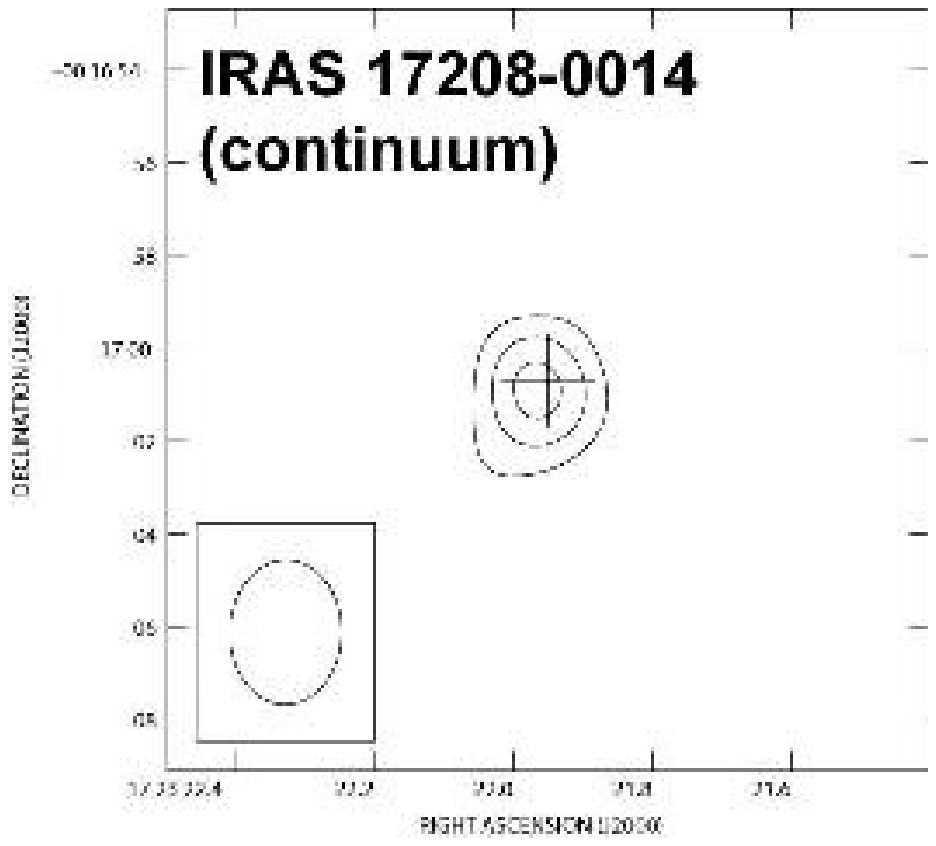} 
\caption{
Continuum maps of UGC 5101, Mrk 273, and IRAS 17208$-$0014.
The crosses show the coordinates of the ULIRG nuclei.
The coordinates in J2000 are
(9$^{h}$35$^{m}$51.61$^{s}$, $+$61$^{\circ}$21$'$11$\farcs$4) for UGC 5101,
(13$^{h}$44$^{m}$42.12$^{s}$, $+$55$^{\circ}$53$'$13$\farcs$5) for
Mrk 273, and
(17$^{h}$23$^{m}$21.95$^{s}$, $-$00$^{\circ}$17$'$00$\farcs$7) for
IRAS 17208$-$0014.
The nuclear coordinates of UGC 5101 are adopted from VLA radio data
\citep{con91}, while those of Mrk 273 and IRAS 17208$-$0014 are based on
the CO peaks in millimeter interferometric maps \citep{dow98}.
The contours are $-$3, 3, 4.5, and 6 mJy beam$^{-1}$ for UGC 5101,
$-$4.5, $-$3, 3, 4.5, and 6 mJy beam$^{-1}$ for Mrk 273, and
$-$4, 4, 6, and 8 mJy beam$^{-1}$ for IRAS 17208$-$0014.
}
\end{figure}

%\clearpage

%---- Figure 2 ----%
\begin{figure}
\includegraphics[angle=0,scale=.8]{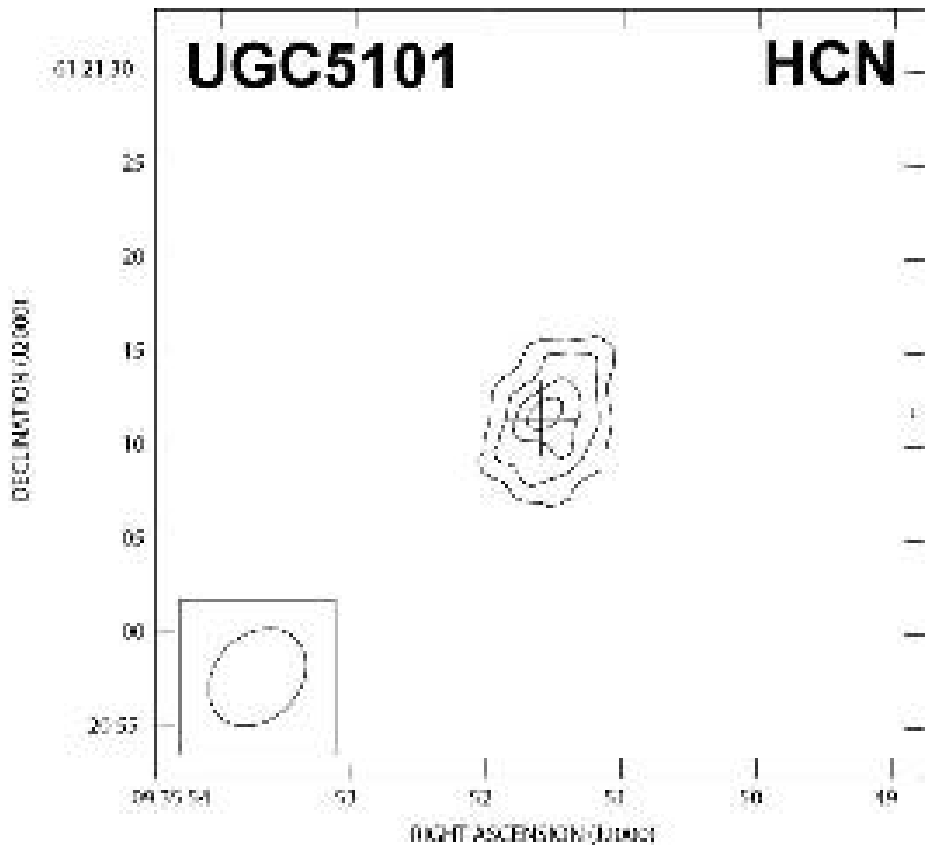} 
\includegraphics[angle=0,scale=.8]{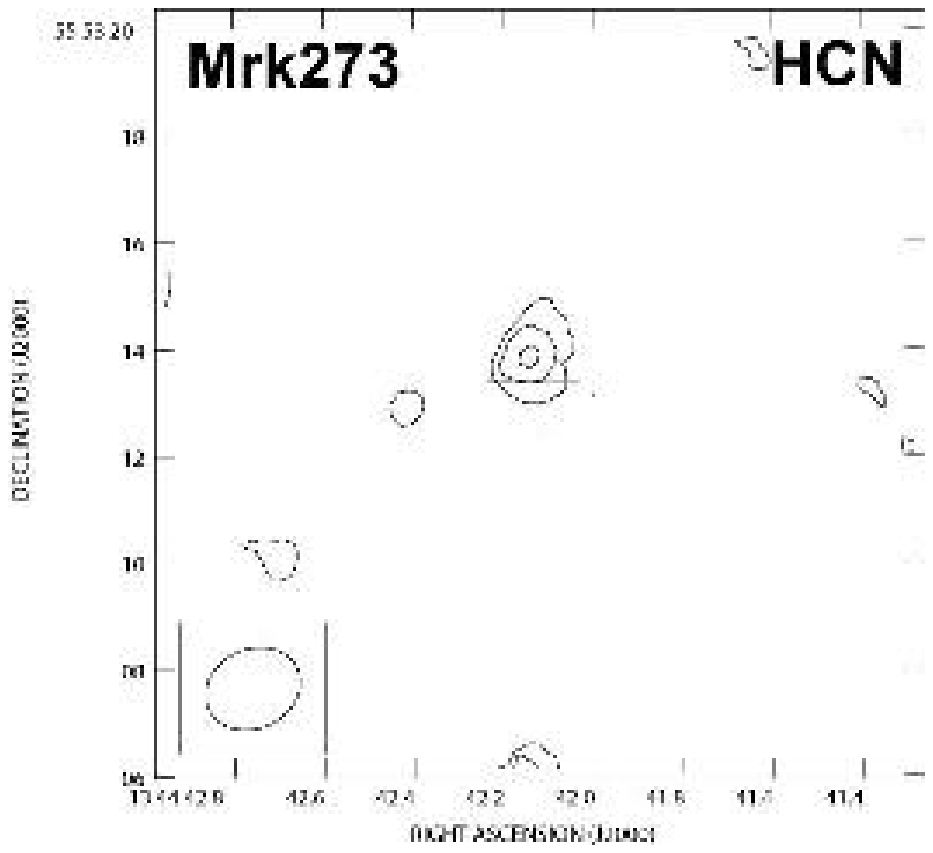}
\includegraphics[angle=0,scale=.8]{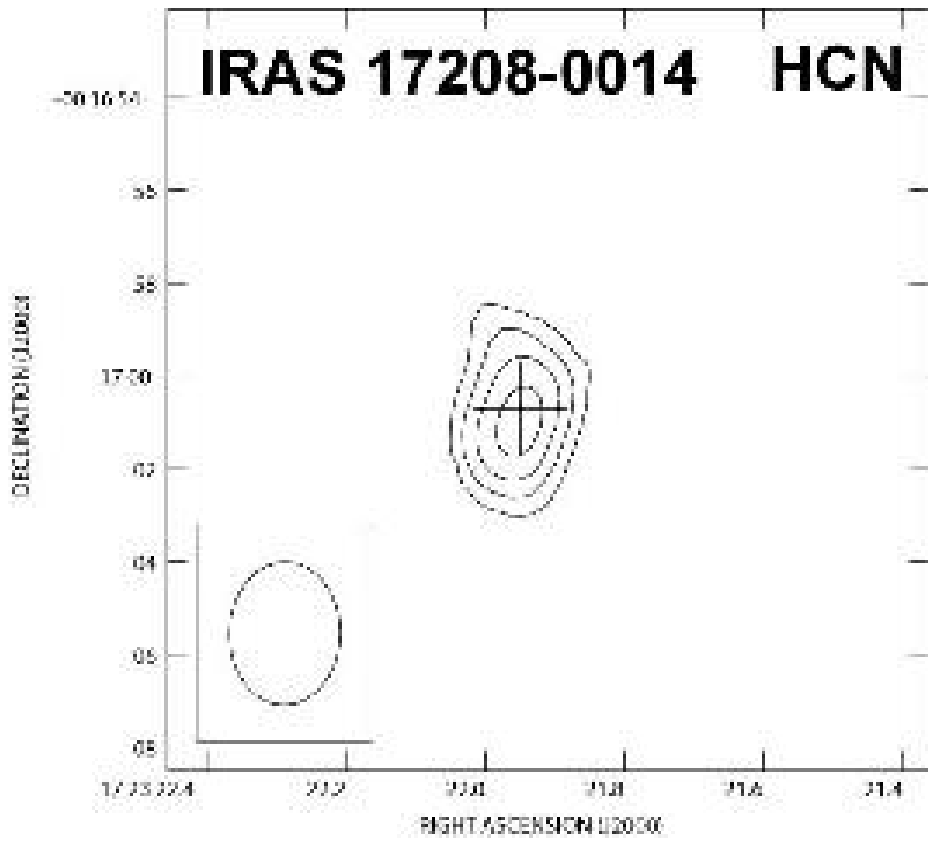} 
\caption{
Integrated intensity map of HCN emission.
The contours are 2.4, 3.6, 4.8, and 6.0 Jy km s$^{-1}$ beam$^{-1}$ for UGC
5101, 2.4, 3.2, and 4.0 Jy km s$^{-1}$ beam$^{-1}$ for Mrk 273, and
4, 6, 8, and 10 Jy km s$^{-1}$ beam$^{-1}$ for IRAS 17208$-$0014.
}
\end{figure}

%---- Figure 3 ----%
\begin{figure}                        
\includegraphics[angle=0,scale=.7]{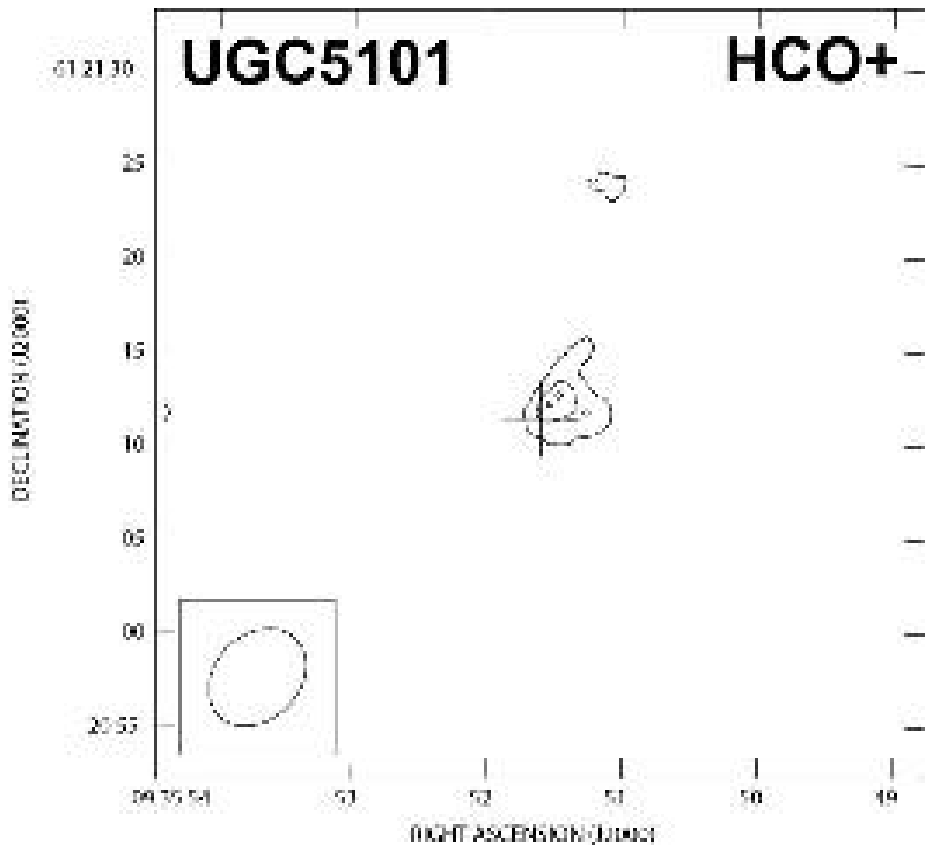} 
\includegraphics[angle=0,scale=.7]{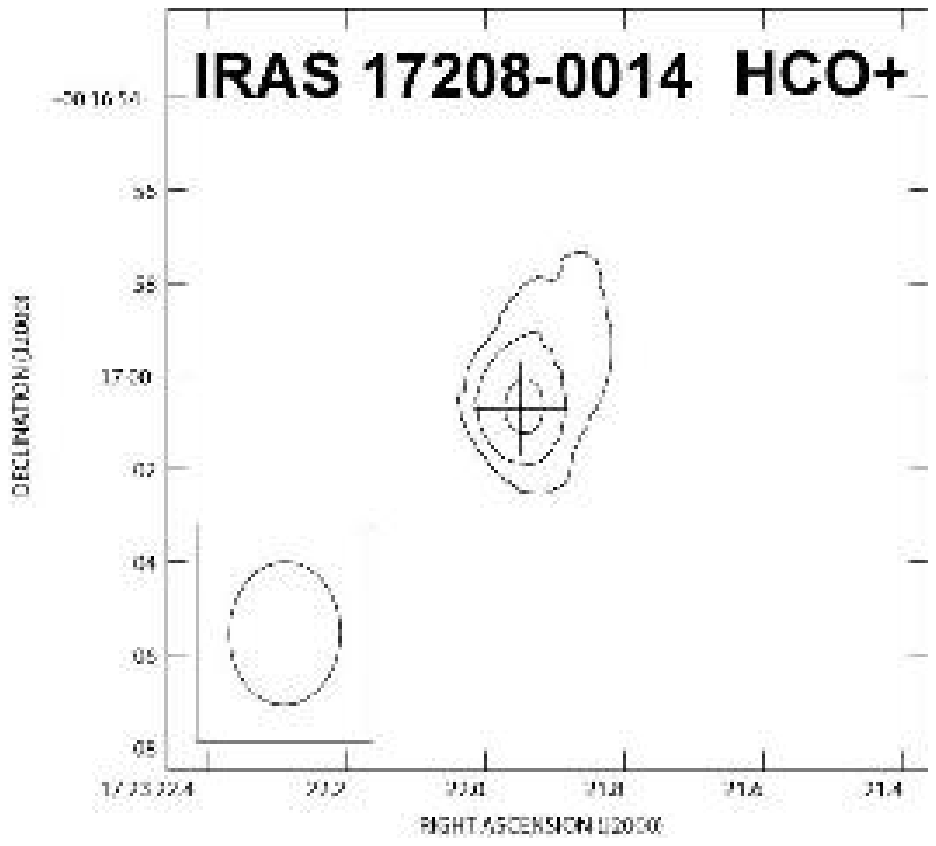} 
\caption{
Integrated intensity map of HCO$^{+}$ emission.
The contours are 1.8, 2.7, and 3.6 Jy km s$^{-1}$ beam$^{-1}$ for UGC 5101, and
2, 4, and 6 Jy km s$^{-1}$ beam$^{-1}$ for IRAS 17208$-$0014.
}
\end{figure}

%---- Figure 4 ----%
\begin{figure}
%\begin{center} {\bf \Huge UGC5101} \\ \vspace{1cm}
%{\bf \Huge HCN \hspace{4cm} HCO+} \end{center}
\includegraphics[angle=0,scale=.54]{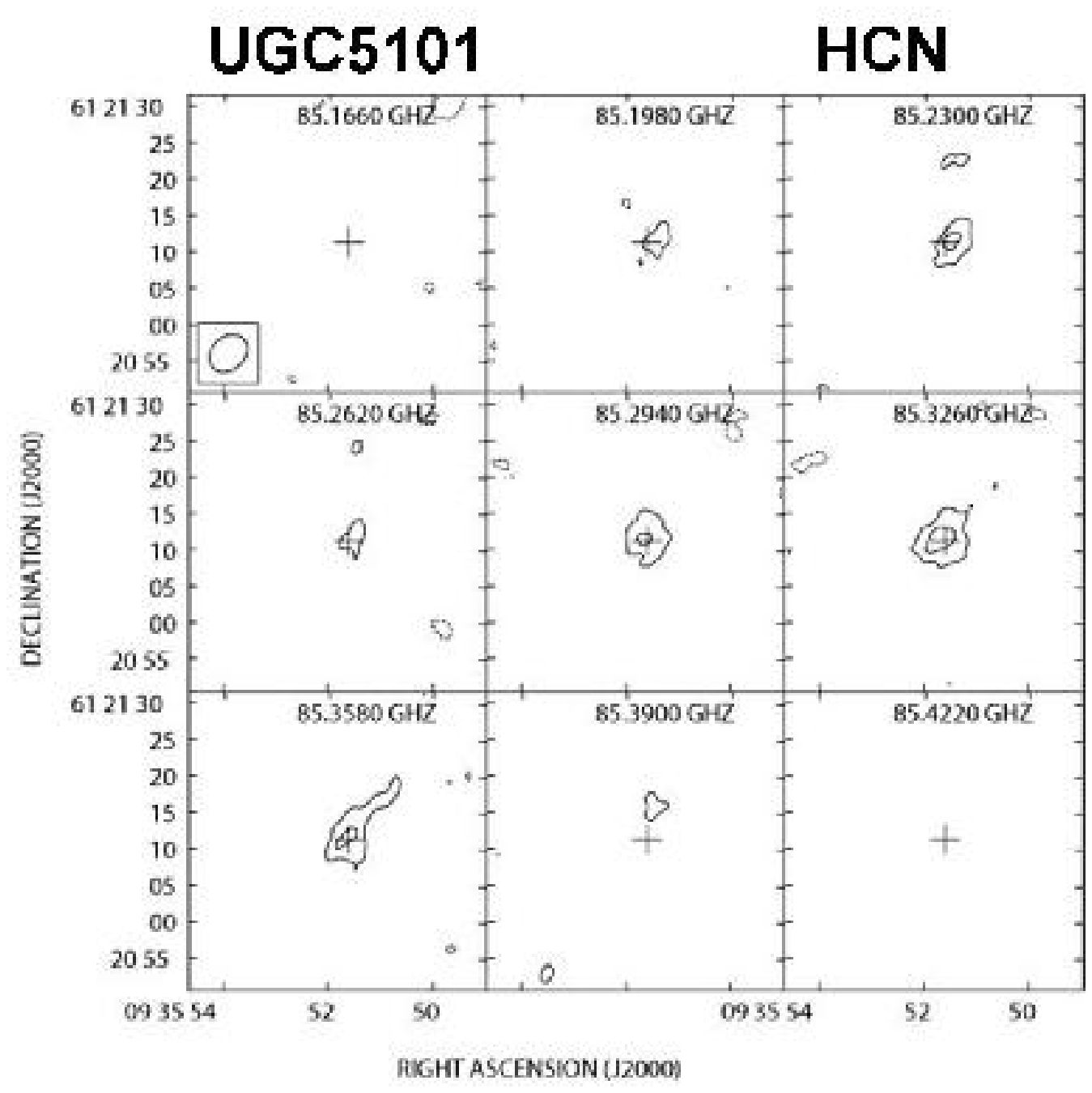}
\includegraphics[angle=0,scale=.54]{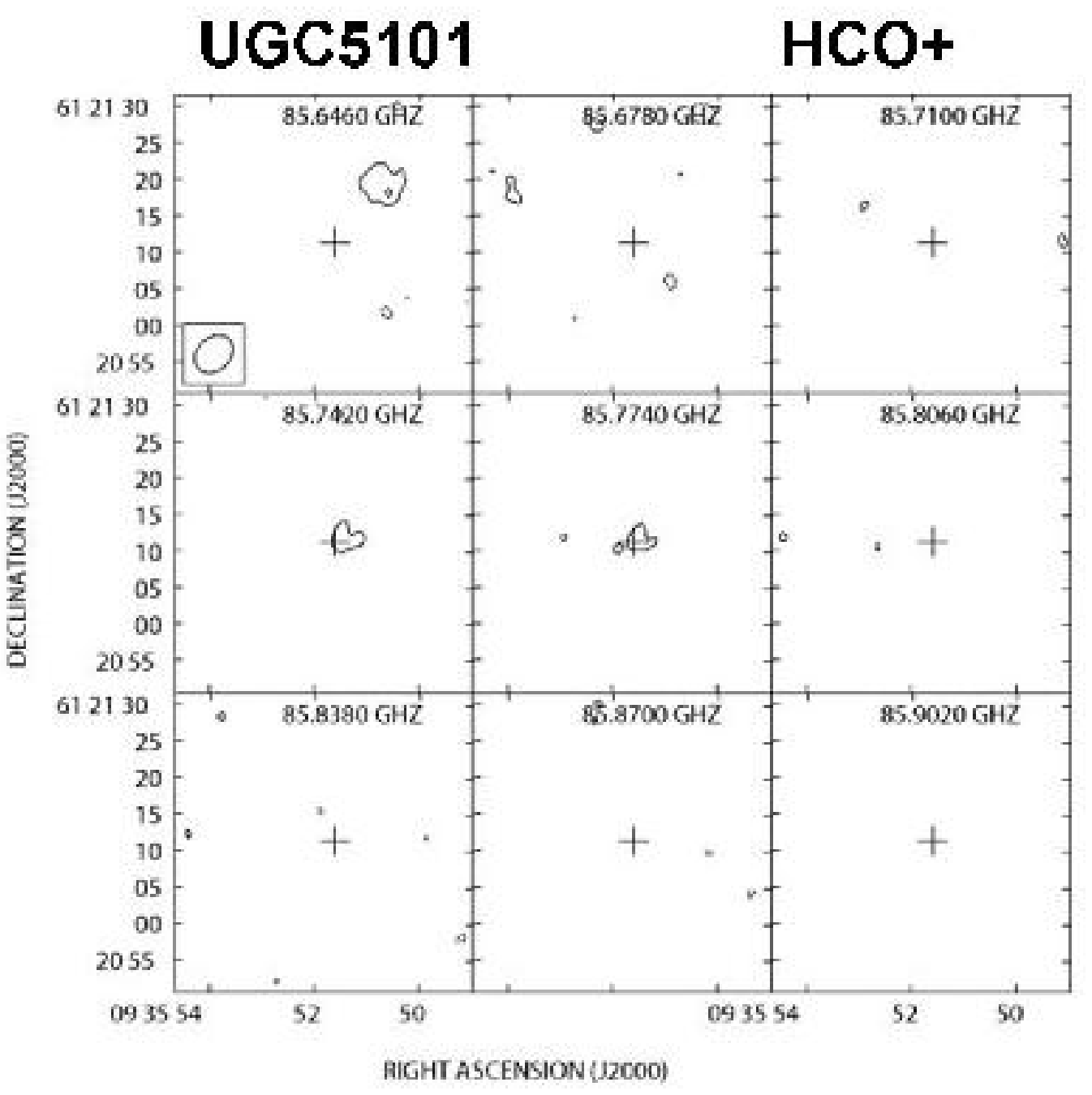}
\caption{
Channel maps of HCN and HCO$^{+}$ emission for UGC 5101.
{\it Left}: HCN emission.
The contours are $-$6, 6, and 10 mJy beam$^{-1}$.
The r.m.s. noise level is $\sim$2.5 mJy beam$^{-1}$.
{\it Right}: HCO$^{+}$ emission.
The contours are $-$6, 6, and 10 mJy beam$^{-1}$.
The r.m.s. noise level is $\sim$2.5 mJy beam$^{-1}$.
}
\end{figure}

%---- Figure 5 ----%
\begin{figure}
%\begin{center} {\bf \Huge IRAS17208-0014} \\ \vspace{1cm}
%{\bf \Huge HCN \hspace{4cm} HCO+} \end{center}
\includegraphics[angle=0,scale=.64]{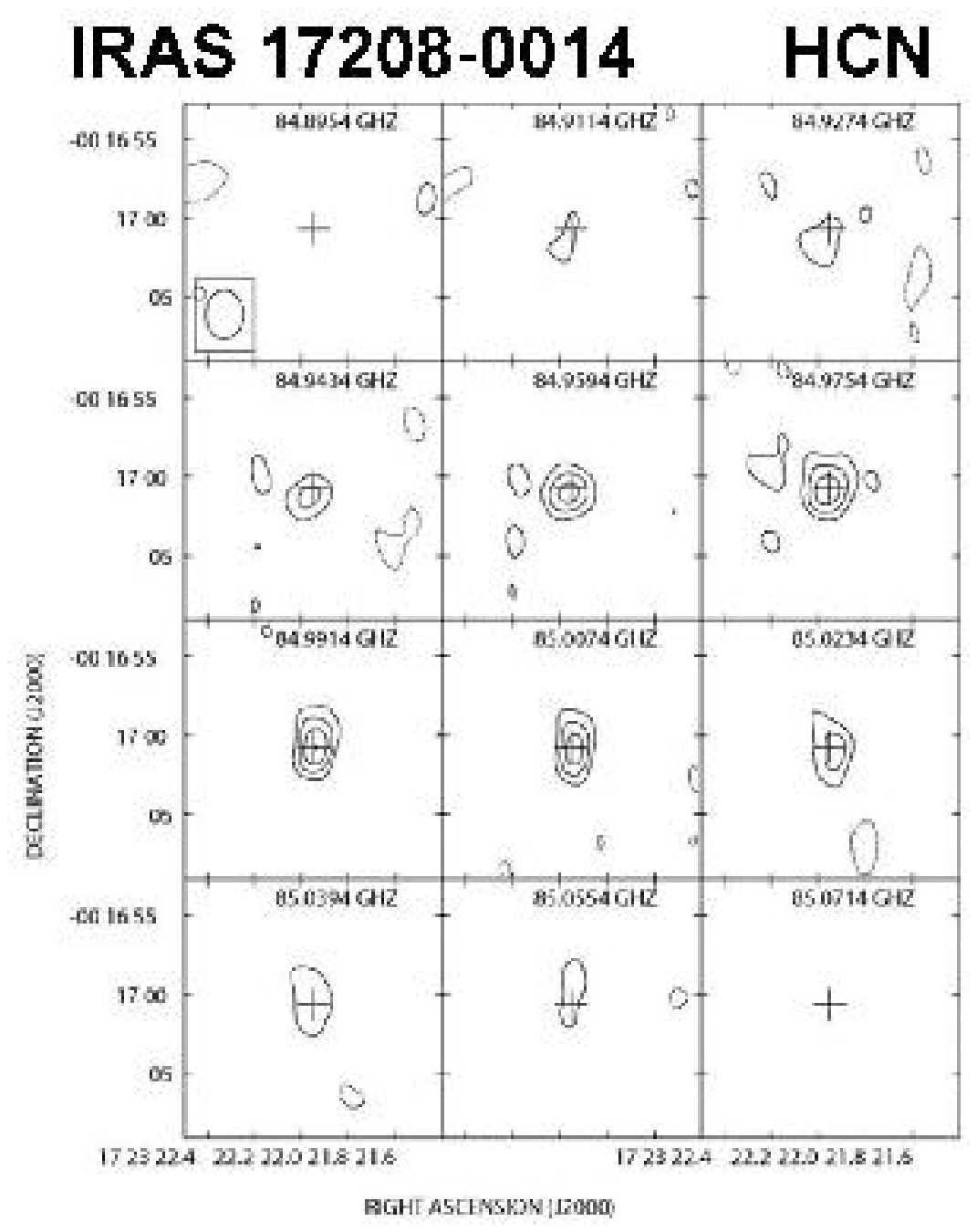}
\includegraphics[angle=0,scale=.64]{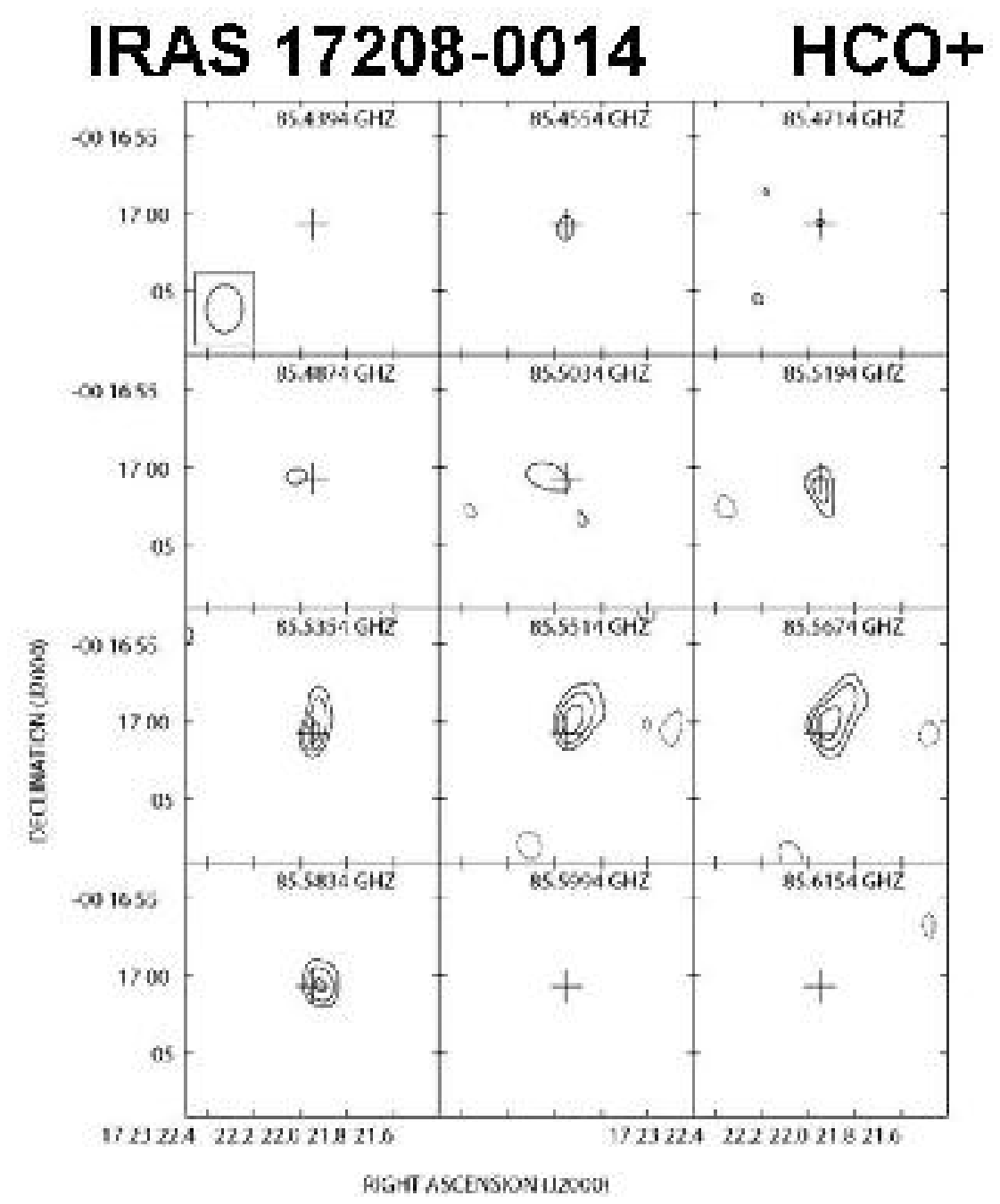}
\caption{
Channel maps of HCN and HCO$^{+}$ emission for IRAS 17208$-$0014.
{\it Left}: HCN emission.
The contours are $-$9, 9, 15, 21, and 27 mJy beam$^{-1}$.
The r.m.s. noise level is $\sim$3 mJy beam$^{-1}$.
{\it Right}: HCO$^{+}$ emission.
The contours are $-$9, 9, 12, and 15 mJy beam$^{-1}$.
The r.m.s. noise level is $\sim$3 mJy beam$^{-1}$.
}
\end{figure}

%---- Figure 6 ----%
\begin{figure}
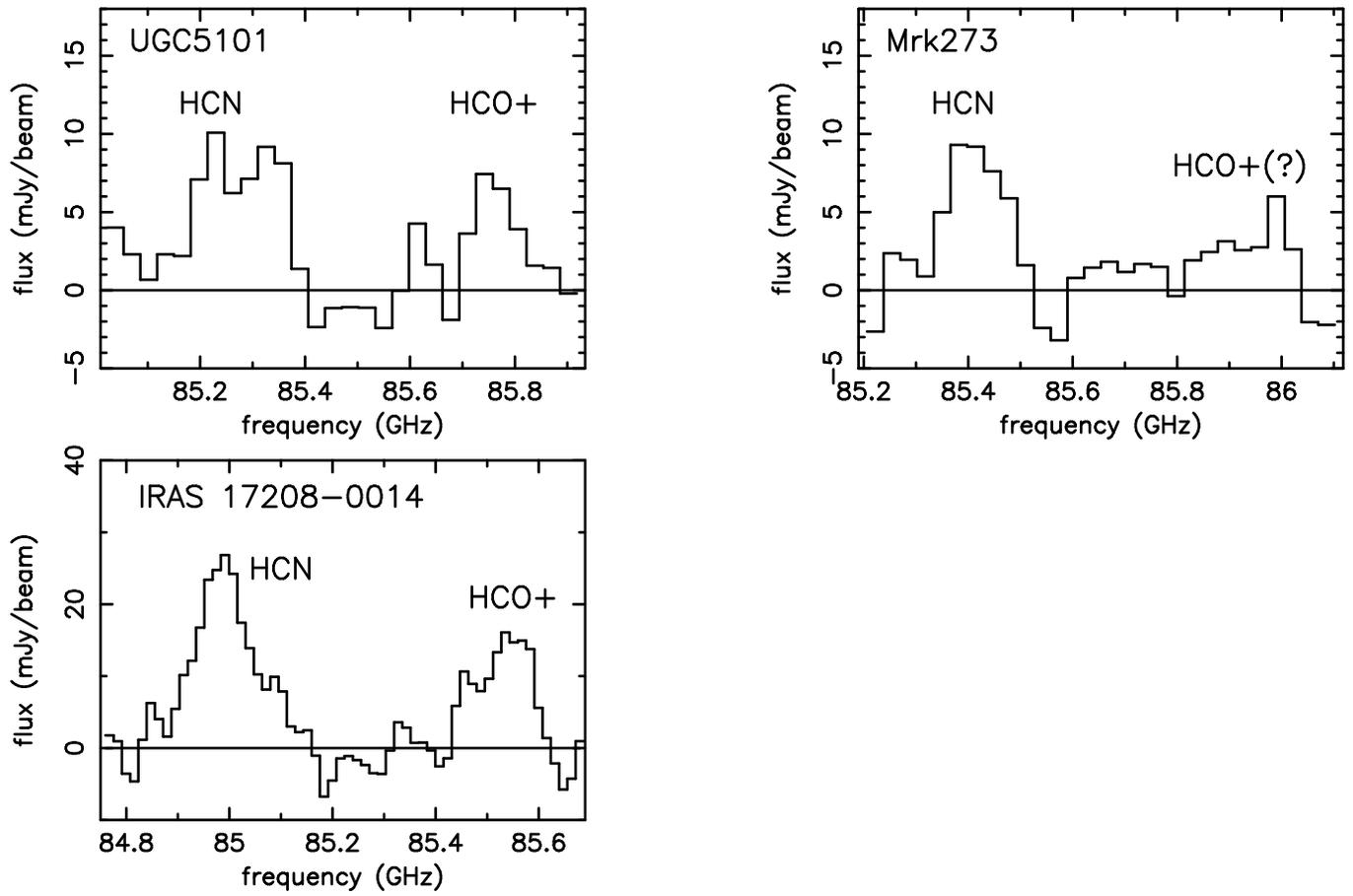

%\begin{center} {\bf \Huge Peak spectrum} \end{center}
\includegraphics[angle=-90,scale=.35]{f6a.eps} 
\includegraphics[angle=-90,scale=.35]{f6b.eps}
\includegraphics[angle=-90,scale=.35]{f6c.eps} \\
\caption{
HCN and HCO$^{+}$ spectra at the nuclei of the ULIRGs.
The abscissa is the observed frequency in GHz and the ordinate is the flux in
mJy beam$^{-1}$.
}
\end{figure}

%---- Figure 7 ----%
\begin{figure}
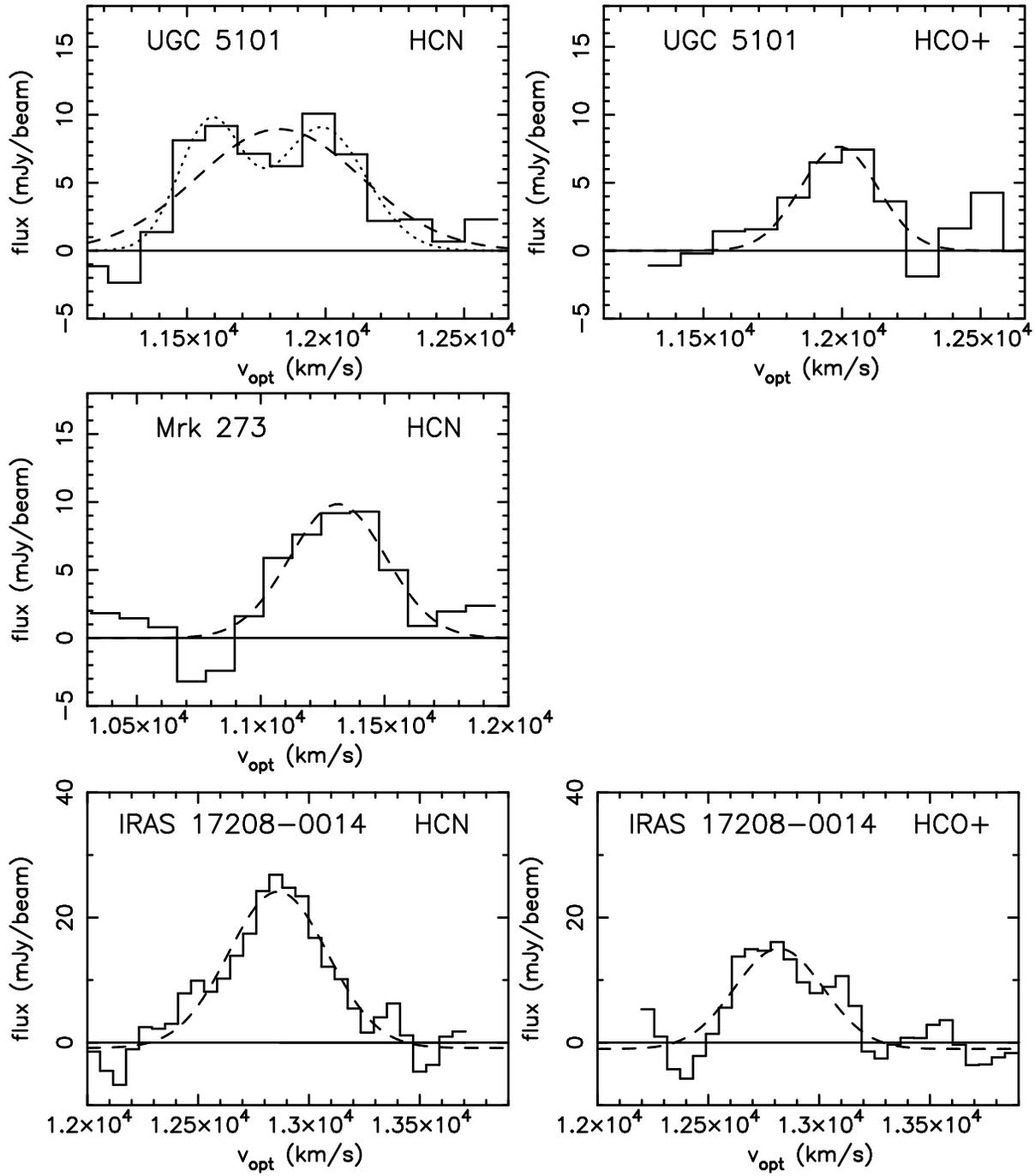

%\begin{center} {\bf \Huge Gaussian fit} \end{center}
\includegraphics[angle=-90,scale=.35]{f7a.eps} 
\includegraphics[angle=-90,scale=.35]{f7b.eps} \\
\includegraphics[angle=-90,scale=.35]{f7c.eps} \\
\includegraphics[angle=-90,scale=.35]{f7d.eps} 
\includegraphics[angle=-90,scale=.35]{f7e.eps} \\
\caption{
Gaussian fits to the HCN and HCO$^{+}$ emission lines.  The
abscissa is the LSR velocity \{v$_{\rm opt}$ $\equiv$
($\frac{\nu_0}{\nu}$ $-$ 1) $\times$ c\} in km s$^{-1}$ and the
ordinate is the flux in mJy beam$^{-1}$.  Single Gaussian fits are
used as defaults and are shown as dashed lines.  For the HCN line of
UGC 5101, since there is evidence for a double peak, a double Gaussian
fit is also attempted and shown as the dotted line.
}
\end{figure}

%---- Figure 8 ----%
\begin{figure}
\begin{center}
\includegraphics[angle=-90,scale=.8]{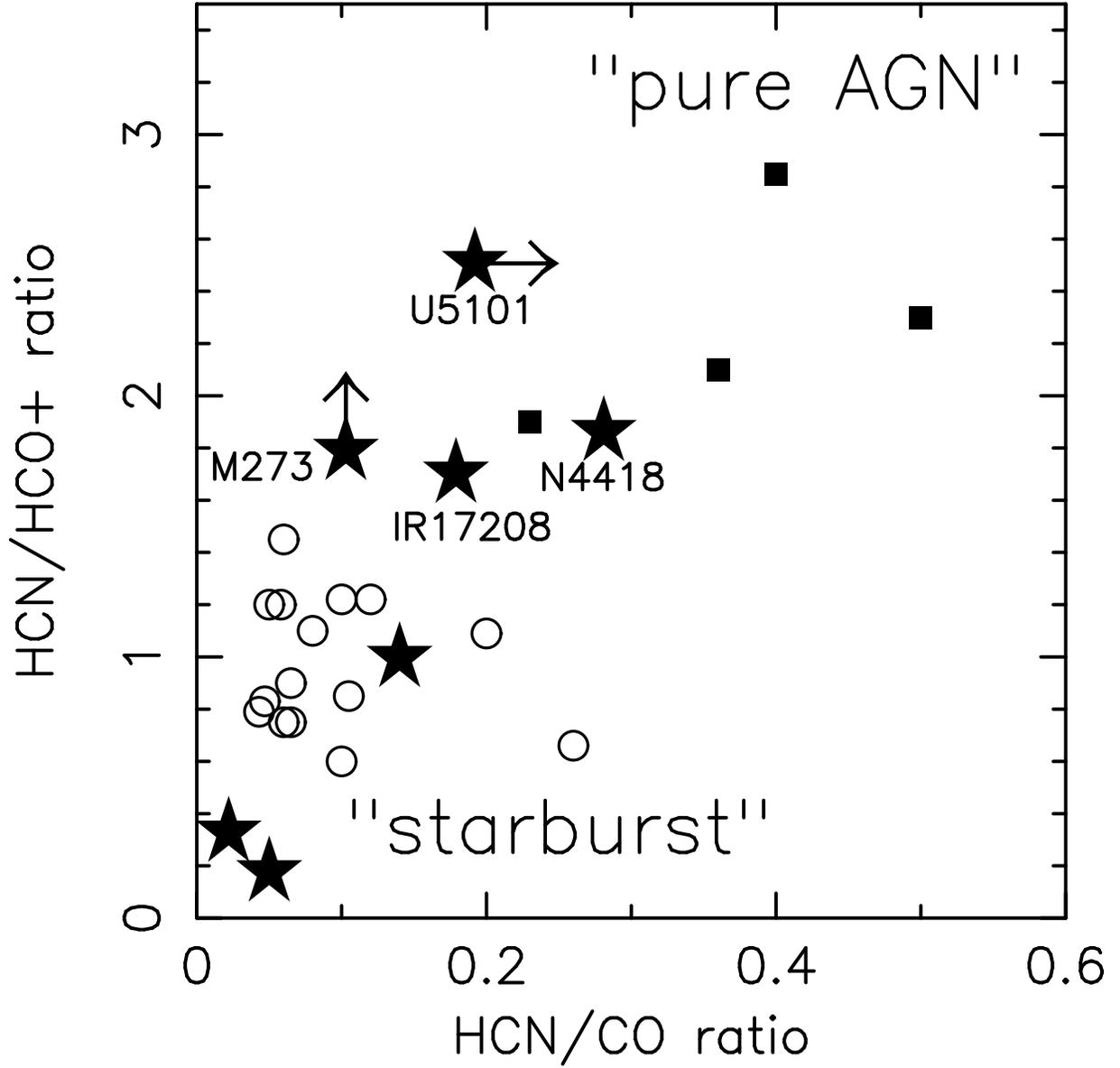}
\end{center}
\caption{
HCN/HCO$^{+}$ (ordinate) and HCN/CO (abscissa) ratios in
brightness temperature ($\propto$ $\lambda^{2}$ $\times$ flux).  UGC
5101, Mrk 273, IRAS 17208$-$0014, and NGC 4418 \citep{ima04} are
plotted as filled stars.  The three main nuclei of Arp 299 (Imanishi
et al. 2006, in preparation) are also plotted as filled stars on the
lower left side.  The HCN/HCO$^{+}$ ratios are derived directly from
our interferometric data, while the HCN/CO ratios are derived by
combining our data with CO data in the literature.  The plot of
NGC4418 has been updated from \citet{ima04} to include new
interferometric CO data by Dale et al.(2005).  Other data points are
taken from \citet{koh01} and \citet{koh05}, where sources with
AGN-like (starburst-like) ratios are marked with filled squares (open
circles).
}
\end{figure}

\end{document}